\documentclass[%
 reprint,
 amsmath,amssymb,
 aps,
]{revtex4-1}

\usepackage{graphicx}
\usepackage{dcolumn}
\usepackage{bm}

\begin{document}
\title{Polymer Welding: Strength Through Entanglements}
\author{Ting Ge$^1$, Flint Pierce$^{2,3}$, Dvora Perahia$^3$, Gary S. Grest$^2$, and Mark O. Robbins$^1$}
\affiliation{$^1$Department of Physics and Astronomy, Johns Hopkins University, Baltimore, MD 21218 USA}
\affiliation{$^2$Sandia National Laboratories, Albuquerque, NM 87185 USA}
\affiliation{$^3$Department of Chemistry, Clemson University, Clemson, SC 29634 USA}

\date{\today}
\begin{abstract}
Large-scale simulations of thermal welding of polymers are performed to investigate the rise of mechanical strength at the polymer-polymer interface with the welding time $t_{\rm w}$. The welding process is in the core of integrating polymeric elements into devices as well as in thermal induced healing of polymers; processes that require development of interfacial strength equal to that of the bulk.  Our simulations show that the interfacial strength  saturates at the bulk shear strength much before polymers diffuse by their radius of gyration.  Along with the strength increase, the dominant failure mode changes from chain pullout at the interface to chain scission as in the bulk.  Formation of sufficient entanglements across the interface, which we track using a Primitive Path Analysis is required to arrest catastrophic chain pullout at the interface.  The bulk response is not fully recovered until the density of entanglements at the interface reaches the bulk value.  Moreover, the increase of interfacial strength before saturation is proportional to the number of interfacial entanglements between chains from opposite sides. 
\end{abstract}
\maketitle

Thermal welding is a common means of joining polymeric elements \cite{jone99, wool95}.
Two polymer surfaces are brought into close contact above their glass transition temperature $T_g$ and are allowed to interdiffuse 
for a welding time $t_{\rm w}$.
Polymer melt dynamics suggests that for homopolymer melts the properties of the weld should be indistinguishable from those of the bulk
once the chains have diffused by about their radius of gyration $R_{\rm g}$.
In practice, however, welds reach the bulk strength at much shorter times \cite{wool95}.
A key question is what determines the rapid rise in interfacial strength
and how it is related
to mass uptake across the surface and polymer entanglement.
Entanglements are topological constraints of polymers by other chains \cite {degennes71} that control their bulk visco-elastic and plastic response.

Experiments have quantified the strength of welds by measuring the interfacial fracture toughness in tensile fracture and the peak shear strength of lap
joints \cite{jud81, kline88, schnell98}.
Both quantities grow as $t_{\rm w}^{1/4}$ at short times and then saturate at their bulk values.
Several different molecular mechanisms have been proposed to explain the scaling of strength with $t_{\rm w}$.
Some assume the strength is simply related to interpenetration depth
\cite{wool81,schnell99,brown01} or the areal density of chain segments bridging the interface \cite{prager81} or the contour length of bridging segments
\cite{degennes89}. 
In many cases these models are motivated by physical pictures of the
development of entanglements at the interface, but chain friction
may also be important \cite{benkoski02}.

It is difficult to distinguish between the proposed strengthening mechanisms with experiments.
Entanglements are not directly observable and
experiments are usually restricted to post-analysis of fracture
surfaces \cite{russell93} or bulk scattering \cite{schnell99},
which does not isolate the response of the failing region.
In contrast, computer simulations provide full spatial resolution
throughout the failure process, allowing macroscopic stresses to be directly
related to molecular structure and dynamics.
Recently developed methods also enable tracking of entanglements on a microscopic level \cite{everaers04, kroger05, tzoumanekas06}.

In this Letter, we present results from large-scale molecular dynamics (MD) simulations of welding between surfaces of highly entangled homopolymers.
The interfacial strength after a welding time $t_{\rm w}$ is determined by a simple shear test that mimics experiments \cite{jud81, kline88, schnell98}.
As in the experiments, the interfacial strength rises linearly with $t_{\rm w}^{1/4}$ before saturating at the bulk value well before the time $t_{\rm d}$ for the chains to diffuse a distance $R_{\rm g}$.
The dominant failure mechanism changes from chain pullout at small $t_{\rm w}$ to chain scission at large $t_{\rm w}$ and in the bulk.
Evolution of entanglements during welding is tracked using a Primitive Path Analysis (PPA) algorithm \cite{everaers04, hoy07b}.
The crossover to bulk response coincides with a rise in the entanglement density at the interface to the bulk density.
Moreover, the areal density of entanglements between chains from opposing
surfaces is linearly related to the interfacial strength at small $t_{\rm w}$.

Our simulations employ a canonical bead-spring model \cite{kremer90}
that captures the properties of linear homopolymers.
Each polymer chain contains $N$ spherical beads of mass $m$.
All beads interact via the truncated shifted Lennard-Jones potential
\begin{equation}
U_{\rm LJ} (r)=4u_0 [(a/r)^{12}-(a/r)^6-
(a/r_{\rm c} )^{12}+(a/r_{\rm c})^6] \ \ ,
\end{equation}
where $r_{\rm c}$ is the cutoff radius and $U_{\rm LJ} (r)=0$ for $r>r_{\rm c}$.
All quantities are expressed in terms of the molecular diameter $a$,
the binding energy $u_0$, and the characteristic time $\tau=a(m/u_0 )^{1/2}$.

For equilibration, beads along the chain were connected by an additional unbreakable finitely extensible nonlinear elastic (FENE) potential
\begin{equation}
U_{\rm FENE} (r)=-\frac{1}{2}kR_0^2 \ln [1-(r/R_0 )^2] \ \ ,
\end{equation}
with $R_0=1.5a$ and $k=30 u_0 a^{-2}$.
For mechanical tests, chain scission plays an essential role and a 
simple quartic potential was used
\begin{equation}
U_{Q} (r)=K(r-R_{\rm c} )^2 (r-R_{\rm c} )(r-R_{\rm c}-B)+U_0 \ \ ,
\end{equation}
with $K=2351u_0/k_B$, $B=-0.7425a$, $R_{\rm c}=1.5a$, and $U_0=92.74467u_0$.
This potential gives the same equilibrium bond length as $U_{\rm FENE}$ and prevents
chains from crossing each other so that entanglements can be studied.
The bonds break at a force that is 100 times higher than the
interchain $U_{\rm LJ}$, consistent with experiments and previous
simulations \cite{rottler02a,stevens01}.
Previous work has shown that the entanglement length for this model is $N_e = 85 \pm 7$
and that the mechanical response for $N=500$ is characteristic of
highly entangled (large $N$) polymers \cite{rottler02a, rottler02b, rottler03, hoy07, hoy08}.

The equations of motion were integrated using a velocity-Verlet algorithm with a time step $\delta t \leq 0.01\tau$.
The temperature was held constant by a Langevin thermostat 
with a damping constant $\Gamma$ \cite{kremer90}.
A million $\tau$ will be abbreviated as $1 M\tau$.
All simulations were carried out using the LAMMPS parallel
MD code \cite{plimpton95}.

Two thin films were constructed following the standard methodology discussed by Auhl et al.~\cite{auhl03}.
Each film contains $M=4800$ chains of length $N=500$ beads or a total 2.4 million beads.
Periodic boundary conditions were applied along the $x$- and $y$- directions with dimensions $L_x=700a$ and $L_y=40a$.
The thickness in the $z$-direction was maintained at $L_z=100a$ using two
repulsive confining walls.
Each film was well equilibrated at a temperature $T=1.0 u_0/k_B$ with
$r_{\rm c}=2.5a$, $\Gamma = 0.1 \tau^{-1}$ and pressure $P=0$ maintained by expansion/contraction along the $x$-direction.
To form the welding interface at $z=0$, the films were placed as close to contact as possible without overlap.
The interdiffusion was along the z-direction.
During interdiffusion, volume was held fixed by repulsive walls perpendicular to the $z$-direction.

After welding for a time $t_{\rm w}$, the system was quenched rapidly below
the glass temperature $T_g \approx 0.35 u_0/k_B$.
First the cutoff radius was reduced to $r_{\rm c}=1.5a$
to decrease computational cost, reduce density changes
and facilitate comparison with past mechanical
studies \cite{rottler02a, rottler02b, rottler03, hoy07, hoy08}.
Then the temperature was quenched at constant volume with a rate $\dot{T} =-10^{-3}  u_0/(k_B \tau)$ to $T=0.5 u_0/k_B$
where $P=0$.
Subsequent quenching to $T=0.2 u_0/k_B$ was done at $\dot{T} = -2 \times 10^{-4} u_0/(k_B \tau)$ and $P=0$.
A Nose-Hoover barostat with time constant
$50 \tau$ was applied to $P_{\rm xx}$ and $P_{\rm yy}$.
The repulsive walls were maintained at $z=\pm L_z$.
We verified that our conclusions are not sensitive to the details of
the quench protocol or geometry.

Shear was applied to the interface in a manner similar to a shear test
of a lap joint
\cite{wool95}.
Beads within $5a$ of the top and bottom were held rigid and displaced
at constant velocity
in opposite directions along the $y-$axis.
The average strain rate in the film,
$d\gamma/dt=2 \times 10^{-4} \tau^{-1}$, was low enough that it did not affect
the mode of failure and stress had time to equilibrate across the
system \cite{rottler03c}.
The shear stress $\sigma$ was determined from the mean lateral force per unit area applied by the top and bottom walls.
The temperature was maintained at $T=0.2 u_0/k_B$ with
a Langevin thermostat ($\Gamma = 1 \tau^{-1}$) acting only on the x-component  to avoid biasing the flow.

\begin{figure}[htb]
\includegraphics[width=0.4\textwidth]{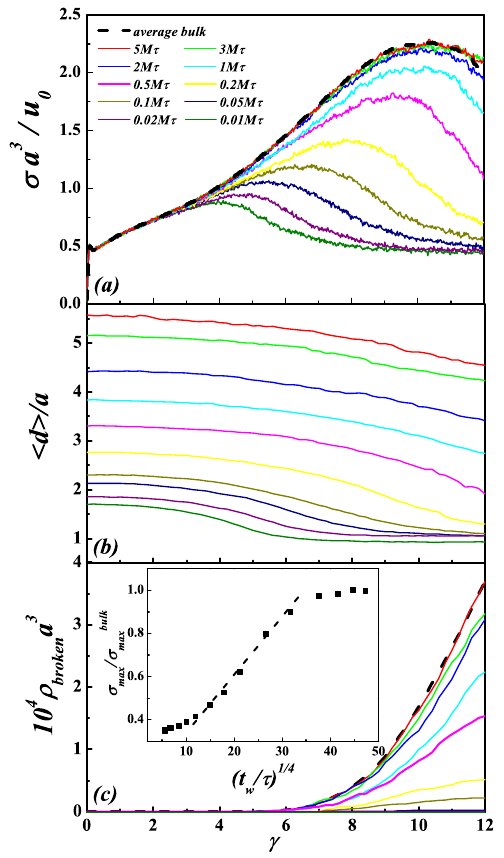}
\caption{
(a) Shear stress $\sigma$ versus shear strain $\gamma$ for
the indicated $t_{\rm w}$.
(b) Average interpenetration depth $\left< d \right>$ versus strain.
(c) Number density $\rho_{\rm broken}$ of broken bonds versus $\gamma$.
The inset shows the maximum shear stress $\sigma_{\rm max}$ 
normalized by its average bulk value
$\sigma_{\rm max}^{\rm bulk}$ versus $t_{\rm w}^{1/4}$. 
}
\label{fig:gamma}
\end{figure}

Stress-strain curves for polymer welds at different $t_{\rm w}$ are compared
to the bulk response in Fig.~\ref{fig:gamma}(a).
The bulk curve is typical of amorphous polymers.
A narrow elastic region is followed by yield and strain hardening - a gradual increase in stress with strain.
At sufficiently large strain, the material begins to fail.
The shear stress reaches a maximum value $\sigma_{\rm max}^{\rm bulk}$
and then drops.

Even for the shortest $t_{\rm w}$, the stress-strain curve follows
the bulk response up to $\gamma \sim 4$.
As $t_{\rm w}$ increases, the stress follows the bulk curve to larger $\gamma$.
For $t_{\rm w} \geq 2 {\rm M}\tau$ the response is nearly indistinguishable from bulk.
In experiments, the strength of the weld is characterized by the maximum
shear stress $\sigma_{\rm max}$ before failure and plotted against $t_{\rm w}^{1/4}$ to
test scaling predictions \cite{wool95}.
Our results, shown in the inset of Fig.~\ref{fig:gamma}(c), are very similar to experiments \cite{wool95}.
There is a constant strength from van der Waals (Lennard-Jones) interactions at short times before appreciable interdiffusion.
The strength saturates to the bulk value as $t_{\rm w}$ increases past
$1 {\rm M}\tau$ and is statistically indistinguishable from bulk behavior by about $3 {\rm M} \tau$.
This is much shorter than the
time for polymers to diffuse by their radius of gyration $t_{\rm d} \gg 10M\tau$ \cite{pierce11}.
At intermediate times there is a linear rise in strength that is consistent with
$\sigma_{\rm max} \sim t_{\rm w}^{1/4}$.
Both simulations and experiments are limited to a factor of $\sim 3$ change in stress
that prevents precise testing of power law scaling.
However the $t_{\rm w}^{1/4}$ scaling is motivated by an assumption that strength is proportional
to the interdiffusion distance.
This is directly tested as described below.

Simulations allow the change in maximum stress to be correlated with changes in failure mechanism and molecular conformation.
Strain leads to stretching and orientation of polymer chains that is
directly related to strain hardening \cite{hoy07, ge10, chen09b}.
Starting near $\gamma \sim 4$ we see a noticeable tension along polymer
backbones that grows rapidly with $\gamma$.
This tension acts to pull back any chain segments that have diffused
across the interface.
One way to quantify chain pullout is to measure the evolution of the average
interpenetration depth $\left< d \right>$ of beads that have crossed
the initial interface ($z=0$)
as a function of strain (Fig. \ref{fig:gamma}(b)).
For $t_{\rm w} < 0.1 {\rm M}\tau$, $\left< d \right>$ begins to drop
at the strain where the stress deviates from the bulk response
and decreases most rapidly near $\sigma_{\rm max}$.
The final interfacial width $\left< d \right>$ is only the size of single bead and corresponds to complete chain pullout.
At this point the stress reaches a constant value that represents
the friction between two separated films.

At larger $t_{\rm w}$, chains cannot be pulled out from the opposing
surface and $\left< d \right>$ decreases by at most a single bead diameter.
The chains have diffused far enough that the
tension required for chain pullout is high enough to break bonds
along the backbone.
Fig.~\ref{fig:gamma}(c) shows the strain dependence of the number density $\rho_{\rm broken}$ of broken bonds averaged over the entire volume between moving layers.
The curves for $t_{\rm w} > 2 {\rm M}\tau$ are similar to the bulk.
Bonds begin to break above $\gamma=6$ and the highest rate of bond breaking is reached near $\gamma=10$.
This point coincides with the peak in shear stress that indicates mechanical
failure.
For $0.1 {\rm M}\tau < t_{\rm w} < 2 {\rm M}\tau$ there is a crossover where the number of broken bonds rises rapidly and the amount of chain pullout measured by the drop in $\left< d \right>$ decreases.
At smaller $t_{\rm w}$, bond breaking is localized near the initial interface.
For $t_{\rm w} > 2 {\rm M}\tau$ bonds break uniformly throughout the system,
confirming that the interface has become as strong as the bulk.
It is interesting to note that a similar transition from chain pullout to
scission occurs with increasing chain length in previous simulations of craze
formation \cite{rottler02a, rottler02b, rottler03}.
Chain scission only occurs when chains are long enough to form entanglements
(typically $N > 2N_e$) that prevent chain pullout.
By analogy, it is natural to expect that the transition to scission and bulk response
occurs when chains have interdiffused enough to form entanglements at the interface.

\begin{figure}[tbh]
\includegraphics[width=0.4\textwidth]{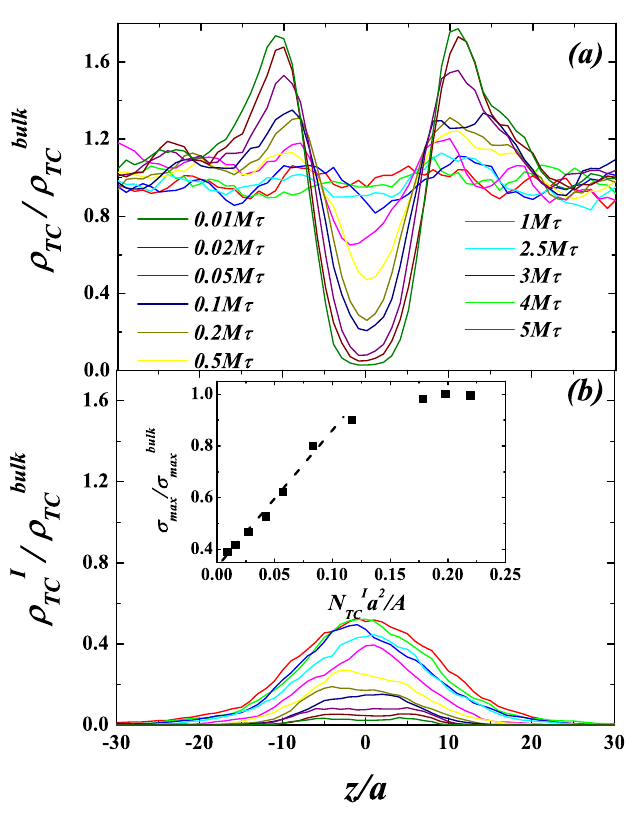}
\caption{
Density profiles for (a) total and (b) interfacial TCs
at the indicated $t_{\rm w}$.
The inset shows the reduced maximum shear stress versus the areal density
of interfacial TCs $N_{\rm TC}^I/A$. 
Error bars are comparable to symbol size.
}
\label{fig:TC}
\end{figure}

Entanglements have proved elusive in experimental studies.
However, the representation of entanglements as binary contacts
between the primitive paths of polymer chains has provided many insights
into the properties of polymer melts \cite{everaers04, kroger05, tzoumanekas06}.
The primitive paths are obtained by fixing the
chain ends and minimizing the chain length without allowing chain crossing. To limit excluded volume effects, 
the chain diameter is then decreased by a factor of $4$
\cite{hoy07b}.
Contacts between the resulting primitive paths are counted to determine
the number of topological constraints (TCs).
We found that the ratio of the density of TCs, $\rho_{\rm TC}$, to the bulk
density, $\rho_{\rm TC}^{\rm bulk}$, was
insensitive to the precise details of the procedure used to identify TCs.
Past bulk studies show that $\rho_{\rm TC}$
is proportional to the entanglement density \cite{everaers04, kroger05, tzoumanekas06, hoy07b} and we 
refer to TCs and entanglements interchangeably below.

Figure \ref{fig:TC}(a) shows the normalized density of TCs as a function of height relative
to the initial interface.
At small $t_{\rm w}$ the chains have not interdiffused enough to produce
any entanglements at the interface, but there are two peaks near $z=\pm 10a$.
These peaks reflect the fact that polymers near free surfaces are
compressed perpendicular to the surface \cite{theodorou88, silberberg88}.
Chains in this pancake-like anisotropic conformation are subject to more TCs.
As welding proceeds, diffusion increases the density of entanglements at the
interface and reduces the peaks on either side.
By $2.5 {\rm M}\tau$ the density has become nearly uniform across the system.

To correlate between $\sigma_{\max}$ and entanglements more
quantitatively, we focus on the number of TCs between chains
that are on opposite sides of the interface at $t_{\rm w}=0$.
All of these formed by interdiffusion and it is natural that these
interfacial entanglements should
be most important in strengthening the interface by preventing
chain pullout.
Fig.~\ref{fig:TC}(b) shows the profiles of the normalized interfacial TC density $\rho_{\rm TC}^I$.
As the polymers diffuse, interfacial TCs spread outward from the interface
and grow in number.
The insert in Fig.~\ref{fig:TC} (b) shows the correlation
between the normalized weld strength $\sigma_{\rm max}/\sigma_{\rm max}^{\rm bulk}$ and the areal density of interfacial TCs,
 $N_{\rm TC}^{\rm I}/A$.
There is a linear correlation between strength and interfacial entanglements at short times.
At long times the interfacial strength saturates while the number of interfacial entanglements continues to grow.

\begin{figure}[htb]
\includegraphics[width=0.4\textwidth]{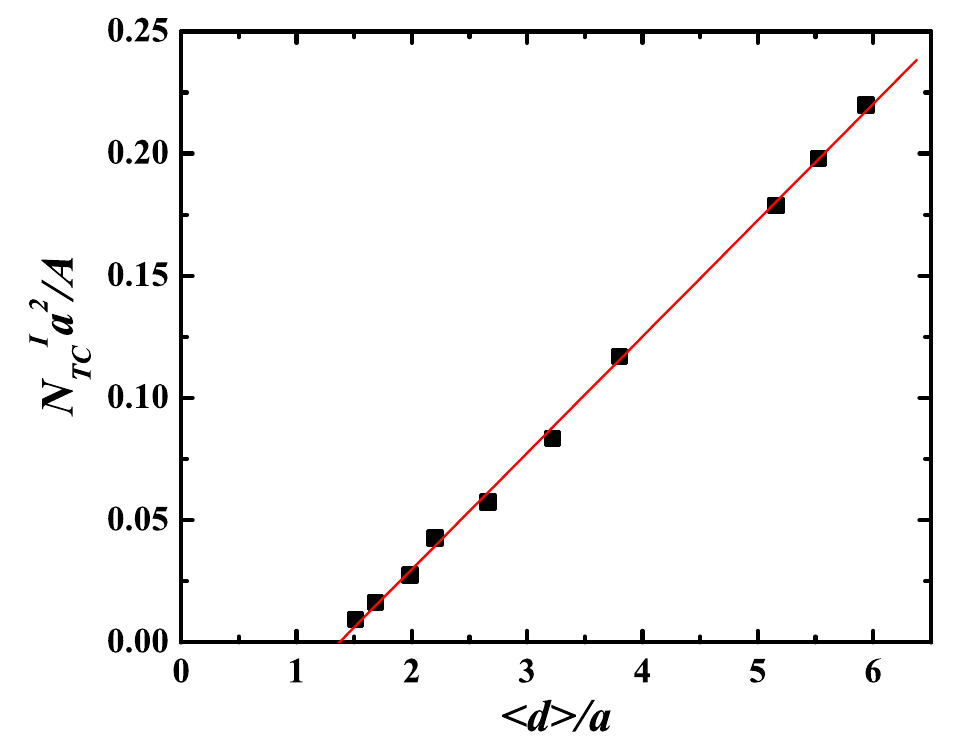}
\caption{
The interfacial density of TCs, $N_{\rm TC}^{\rm I}/A$ versus the average interpenetration depth $\left< d \right>$ and a linear fit.
}
\label{fig:fig3}
\end{figure} 

From the chain-packing model of entanglements \cite{lin87, fetters94, brown96}, the number of interfacial entanglements $N_{\rm TC}^{\rm I}$ scales with the volume spanned by chains that have diffused across the interface.
This can be estimated as $2A\left< d \right>$.
Figure 3 confirms that there is a linear relation between $N_{\rm TC}^{\rm I}/A$ and $\left< d \right>$ over the entire range studied 
\cite{foot1}.
As noted above, welding models have generally assumed that $\sigma_{max}$ rose linearly with $\left< d \right>$ and then used reptation theory to argue 
that both scale as $t_{\rm w}^{1/4}$.
Subsequent work has shown that interfacial diffusion is more complicated because of the anisotropy that produces
the peaks in entanglement density noted above and the prevalence of chain ends at the interface \cite{wool95,pierce11}.
Indeed, recent simulations show that there is not a single simple scaling exponent over the range of times studied here \cite{pierce11}.
Figures 2 and 3 show that whatever the time dependence is, the fundamental factor determining strength is the entanglement density.
One of the main differences between welding models is that some have assumed a simple proportionality between $N_{\rm TC}^{\rm I}/A$ and $\left< d \right>$
\cite{wool95,benkoski02,prager81,wool81},
while others assume a minimum interpenetration distance is needed for entanglements \cite{brown01,adolf,mikos}.
Fig. 3 supports the latter interpretation, with a minimum distance $\sim 1.5 a$ that is about a third of the distance needed for saturation.

To summarize, we have demonstrated that the development of interfacial strength during welding is closely related to the formation of entanglements across the interface.
The interface becomes mechanical indistinguishable from surrounding regions when the bulk entanglement density is recovered at the interface.
There are then sufficient entanglements to prevent chain pullout at the interface and the joint fails through the bulk mechanism of chain scission.
Before the bulk strength is recovered, the interfacial strength rises linearly with the areal density of interfacial entanglements.
This quantity is not accessible to experiments, but is linearly related to the interdiffusion distance
which has been measured.
These findings should help further development of theoretical descriptions of entanglement evolution across a polymer-polymer interface and constitutive molecular modeling of fracture in polymers. 
Of particular interest will be studies that vary the entanglement density by changing the entanglement length or making systems immiscible.

This work was supported by the National Science Foundation
under grants DMR-1006805, CMMI-0923018, OCI-0963185 and DMR-0907390.
MOR acknowledges support from the Simons Foundation.
This research used resources at the National Energy Research Scientific Computing Center (NERSC), which is supported by the Office of Science of the United States Department of Energy under Contract No. DE-AC02-05CH11231. Research was carried out in part, at the Center for Integrated Nanotechnologies, a U.S. Department of
Energy, Office of Basic Energy Sciences user facility. Sandia National Laboratories is a multi-program laboratory managed and operated by Sandia Corporation, a wholly owned subsidiary of Lockheed Martin Corporation, for the U.S. D/Gpartment of Energy's National Nuclear Security Administration under contract DE-AC04-94AL85000.

\begin{thebibliography}{38}
\bibitem[{foo()}]{foot1}
\bibinfo{note}{To reduce the effects due to capillary waves at the interface,
  the width was measured from the local midplane where the densities for beads
  from opposite sides are equal, rather than simply from $z=0$.}
\end{thebibliography}

\end{document}